# Magnetic field induced strong valley polarization in the three dimensional topological semimetal LaBi

Nitesh Kumar[1*], Chandra Shekhar[1], J. Klotz[2,3], J. Wosnitza[2,3] and Claudia Felser[1*]

[1]Max Planck Institute for Chemical Physics of Solids, 01187 Dresden, Germany

[2]Hochfeld-Magnetlabor Dresden (HLD-EMFL), Helmholtz-Zentrum Dresden-Rossendorf, 01328 Dresden, Germany

[3]Institut für Festkörperphysik, TU Dresden, 01062 Dresden, Germany

## Abstract

*LaBi is a three-dimensional rock-salt-type material with a surprisingly quasi-two-dimensional electronic structure. It exhibits excellent electronic properties such as the existence of nontrivial Dirac cones, extremely large magnetoresistance, and high charge-carrier mobility. The cigar-shaped electron valleys make the charge transport highly anisotropic when the magnetic field is varied from one crystallographic axis to another. We show that the electrons can be polarized effectively in these electron valleys under a rotating magnetic field. We achieved a polarization of 60 % at 2 K despite the co-existence of three-dimensional hole pockets. The valley polarization in LaBi is compared to the sister compound LaSb where it is found to be smaller. The performance of LaBi is comparable to the highly efficient bismuth.*

The search of an extra degree of freedom has played a significant role in modern technological advancements. In spintronics, one adds the spin degree of freedom in charge devices [1,2], whereas in the magnetoelectric multiferroics, the polarization can be manipulated by a magnetic field and *vice versa* [3,4]. In recent years, a new field named valleytronics shows great promise particularly in materials with two-dimensional (2D) electronic structures based on their layered crystal structure. For example in graphene, linearly dispersing conical bands (valleys) touch at the high symmetry points of the Brillouin zone (BZ) and one of the exciting aspects for exploiting such a band structure is to occupy (polarize) one valley much larger than another in a controlled fashion. This enables graphene to be used for valley valves and valley filters [5,6]. In contrast to graphene, monolayers of some of the transition metal dichalcogenides (TMDs) inherit broken inversion symmetry with finite band gap. The absence of inversion symmetry and a finite band gap provide crucial electrical and optical control of the valley degree of freedom.[7-9]

An entirely different material class can be envisaged if one considers selectively populating valleys by rotating the material under magnetic field [10]. This requires the material to contain highly anisotropic Fermi surfaces to act as valleys. For example, semimetallic bismuth hosts three cigar-like electron valleys in momentum space. A valley polarization as large as 80 % could be achieved by rotating a single crystal of Bi along its trigonal axis [10]. Despite the huge potential to control the valley polarization in this way, not many materials have been studied. The only other material known in literature is layered $SrMnBi_2$ which shows a polarization of its anisotropic quasi-2D Dirac valleys under crystal rotation [11].

The binary compound LaBi crystallizes in a rock-salt-type structure (Fig. 1(a)) and its anisotropic Fermi surface has long been known [12]. Due to the prediction of a non-trivial band structure in LaBi and other lanthanum mono-pnictides [13] there is strongly revived interest in their electronic properties [14-18]. We have shown earlier the realization of quasi-2D transport and excellent magnetoresistance (MR) performance, which could further be enhanced by tuning the magnetic field direction as a result of the large anisotropy in the Fermi surface [14]. Later, we also found by use of angle resolved photoelectron spectroscopy (ARPES) that LaBi hosts $Z_2$ topology and multiple Dirac points in the first BZ [19].

Most of the research in this class of compounds is limited to investigating the origin of large MR and the ARPES measurements for the presence of two dimensional Dirac cone. However, despite the highly anisotropic MR due to Fermi surface topology, direction dependent transport properties are still scarce in the literature. In this article, we demonstrate that LaBi, because of elliptical Fermi surface with large in-plane to out-of-plane effective mass ratios, is an excellent candidate for valley polarization of electrons. We could achieve a polarization value as large as 60 % by rotating the single crystal under applied magnetic field. In contrast, LaSb with a similar Fermi-surface anisotropy and even higher effective mass ratio shows smaller polarization due. We further point out that the electron-like Fermi surface of LaBi highly resembles that of diamond which has previously been shown to be very effective in generating valley polarized electrons [21].

LaBi is a material with many exotic properties. (1) It exhibits a large mobility (~1×10$^4$ cm$^2$/Vs) and huge MR of the order of 10$^5$ % at 2 K and 9 T because of charge-carrier compensation and orbital texturing [14,16]. (2) It features several Dirac points and topological surface states. (3) At high pressure, it becomes superconducting [22]. LaBi presents an intriguing electronic structure (Fig. 1(b)) where two hole-like bands cross the Femi level near the Γ-point and an electron-like band near the X-point. The electron band forms a highly anisotropic ellipsoidal Fermi pocket in momentum space. Owing to the $Fm\bar{3}m$ crystal symmetry, there are in total six of such pockets in the ±$k_x$, ±$k_y$ and ±$k_z$ directions. Angular-dependent Shubnikov-de Haas (SdH) oscillations allowed us to track the shape of the Fermi pockets. In Fig. 1(d), we show the extremal area of the electron-like Fermi surface which reflects its ellipsoidal shape. The extremal areas are calculated from fast Fourier transforms of the quantum oscillations (Fig. S1) according to the Onsager relation: $F = (\Phi_0/2\pi^2)A_k$ where, $\Phi_0$ is the flux quantum (2.07 x 10$^{-15}$ Tm$^2$) and $A_k$ is the Fermi surface area normal to the applied magnetic field. A detailed SdH study of LaBi can be found in ref. [14].

The anisotropy in the electronic band structure is directly reflected in the transport properties. In order to capture the effect of valleys located along the crystallographic axes, we have measured the resistivity of LaBi with current along [100] and rotated the crystal such that the angle between the current and the magnetic field is always 90º (Fig. 1(c)). This configuration guarantees that the effect of the Lorentz force is constant throughout the measurement. Fig. 2 shows the resistivity of LaBi at 2 K when the crystal is rotated by 2π at different applied magnetic fields. An oscillatory pattern can be resolved at fields as low as 0.3 T. As the field is increased, the oscillations can be observed more easily. At 0º, the resistivity is minimal, increases continuously until 45º and then decreases again, so that the minima and maxima repeat every 90º. This is consistent with the fourfold symmetry of LaBi with respect to the [010] axis. The overall increase in the resistivity at higher fields is a consequence of the large MR in LaBi. The small ripples on top of the wave patterns in the 7 and 9 T data are due to SdH oscillations. At 9 T, the wave-like pattern can easily be observed until 200 K (See Fig. S2). The polar plot of the conductivity at 2 and 100 K in 9 T is shown in Figs. 2b and 2c, respectively.

LaBi is a highly compensated semimetal, which implies that the off-diagonal terms of the resistivity tensor are small compared to the diagonal terms. Hence, the relation $\sigma_{xx} = 1/\rho_{xx}$ is an appropriate approximation. The total conductivity of LaBi has contributions from two hole pockets centered at the Γ-point and three electron pockets centered at the X-points. A smaller, spherical hole pocket is enclosed by a larger star-shaped hole pocket (Fig. 1(b)). The electron pockets are elongated ellipsoids and lead to a pseudo-two-dimensional behavior. The two hole pockets can be considered to be three dimensional compared to the ellipsoidal electron pockets which we call electron valleys. Hence,

$$\sigma_{xx} = \sigma_h + \sum_{i=1-3} \sigma_{ei}. \qquad (1)$$

However, when the magnetic field is rotated in the y-z plane the electron valleys aligned along $k_x$ will virtually behave as isotropic pockets. Therefore, the above equation can be transformed into:

$$\sigma_{xx} = \sigma_h + \sigma_{e(xx)} + \sum_{i=1,2} \sigma_{ei} = \sigma_h + \sigma_{e(xx)} + \sigma_{e1} + \sigma_{e2} \qquad (2)$$

Here, $\sigma_{e1}$ and $\sigma_{e2}$ are conductivity components from the valleys located at $k_y$ and $k_z$, respectively. The first two terms correspond to the 3D and the last two to the quasi-2D conductivity components. As suggested by Zhu *et al.* [10], the best fit to the total conductivity is obtained by using:

$$\sigma_{xx} = \sigma_{3D} + \sum_{i=1,2} \frac{\sigma_{2D}}{1 + r\cos^2\left(\phi + (i-1)\frac{\pi}{2}\right)}$$

$$= \sigma_{3D} + \frac{\sigma_{2D}}{1 + r\cos^2\phi} + \frac{\sigma_{2D}}{1 + r\cos^2\left(\phi + \frac{\pi}{2}\right)}. \qquad (3)$$

Here, *r* is a fit parameter that provides a measure of the conductivity anisotropy due to the elongated Fermi pockets. It incorporates both, anisotropies in effective masses and scattering rates [11]. When the crystal is rotated in the *y-z* plane, an electron pocket is encountered at every π/2. Fig. 3(a) shows the measured $\phi$-dependent conductivity (open circles) at 2 K and 9 T, a fit (red line) using the above equation, individual conductivity components from the electron valleys along $k_y$ (green line) and $k_z$, (blue line), and the 3D component (orange dotted line) comprising the contributions from the hole pockets and electron valleys along $k_x$. The ratio of the conductivity from the ellipsoidal electron valleys to the overall 3D conductivity ($2\sigma_{2D}/\sigma_{3D}$) is 3.35, which is more than 40 times larger than in the two-dimensional material SrMnBi$_2$ [11]. The rather small contribution from the 3D pockets in LaBi is responsible for a valley polarization as large as 60 % as can be seen in Fig. 3(b). A maximum polarization of 80 % was achieved in bismuth having a larger mass anisotropy. The reduced polarization in LaBi may be due to the fact that the 3D component of the total conductivity is composed of two hole pockets and one electron pocket compared to only one spherical hole pocket in bismuth. Interestingly, the potentially highly suitable candidate for valley polarization diamond [21] has a strikingly similar band structure. Elongated ellipsoidal conduction bands are oriented along the {100} axes in diamond, equivalent to the electron valleys in LaBi. However, the additional 3D hole pockets situated at the center of the BZ of LaBi play a very important role in imparting a very large magnetoresistance as a result of carrier compensation.

Fig. 4(a) shows the field-dependent anisotropy factor *r*, at 2 and 30 K. At 2 K, *r* reaches a saturation value of around 8. At 30 K, the anisotropy is smaller than at 2 K in the whole field range. The corresponding fits are shown in Figs. S3 and S4. The anisotropy factor significantly affects the resistivity when the magnetic field is applied along the different crystallographic directions. We observed that the resistance is minimal when the field is aligned along the crystallographic axes (*e.g.* $\phi$ = 0°) and maximal when in between (*e.g.* $\phi$ = 45°) and it. The resistivity as a function of field for $\phi$ = 0° and 45° is shown in Fig. 4(b). The extent of the field-induced resistivity increase at 45° compared to 0° should be proportional to the anisotropy factor *r*. To prove this, we plot in Fig. 4(c) the percentage increase in the resistivity at 45° compared to 0° as a function of field at 2 K and call it $\Delta\rho(\pi/4)$. We find that this quantity has exactly the same field dependence as the anisotropy factor (see Fig. 4(a) for 2 and 30 K). As Fig. 4(c) shows, the valley polarization survives even at 100 K in 9 T. The field-dependent resistivity at 0° and 45° for different temperatures is shown in Fig. S5. This also provides a simpler experimental approach to the anisotropic transport originating from valley polarization as compared to the anisotropy factor *r* in equation (3), because the latter requires a fit to the angle-dependent resistivity data for each field. The temperature and field dependence of *r* in LaBi is quite different from that in both Bi and SrMnBi$_2$. In case of Bi, $r$ is independent of magnetic field up to 40 K after which $r$ increases with field. This means that the magnetic field and hence the orbital magnetoresistance in LaBi acts as a valley valve up to 100 K, in contrast to Bi where it is only true above 40 K [10]. This is also the case for SrMnBi$_2$ except that $r$ does not saturate even at very high field [11].

Importantly, we do not observe any nematic order in LaBi and the crystal rotational symmetry is maintained up to 9 T. In many multi-valley systems, the inherent rotational symmetry from the crystal structure is lifted due to Coulomb interactions [23]. The observation of such order in bulk measurements would require very high-quality crystals. Disorder in the crystals may result in short range order and domain formations and hence requires local probes such as STM [24]. Another possibility to see the lifting of degeneracy is by carrying out measurements at large fields. However, the large amplitudes of the SdH oscillations at high fields [14] and slight crystal misalignments would further make it difficult to observe any nematic ordering.

A very powerful way to confirm the valley polarization would be to investigate the angle-dependent quantum oscillations in more detail. The depopulation of a particular electron valley would result in the vanishing of the corresponding oscillation. However, in LaBi the quantum-oscillation signal of the electron valley disappears because of the intrinsic shape of the valley rather than the valley polarization. This was confirmed by angle-dependent torque measurements (18 T), where we also observe the disappearance of oscillations from the electron valleys at high angles (see Fig. S6). Since the torque measurements do not involve any electric field, one can exclude the effects of valley polarization. This has its origin in the large curvature of the ellipsoidal Fermi surfaces at high angles, which is known to suppress the quantum-oscillation amplitude. The same behavior is also seen in LaSb, a case where we observe significantly smaller valley polarization.

We compare the results of the valley polarization in LaBi with that of the other member of the series, LaSb. The transport properties of both compounds are quite similar with extremely large MR and high-mobility charge carriers. LaSb also contains cigar-like electron pockets with large mass anisotropy, larger than in LaBi [12] (Fig. 5(b)). Despite all these attributes, LaSb exhibits a lower valley polarization compared to LaBi (Fig. 5(a)). As discussed earlier, the valley polarization is defined as $\Sigma\sigma_e/\sigma_{total}$, the smaller value of $\Sigma\sigma_e$ and a large value of $\sigma_{total}$ would imply a smaller valley polarization in LaSb. Due to a larger magnetoresistance effect in LaSb compared to LaBi, $\sigma_{total}$ at a particular magnetic field is larger in LaSb. For both LaSb and LaBi, the $m^*$ (effective mass) of hole pockets are similar, but the $m^*_{perp}$ of electron pocket in LaSb (= $0.13m_0$ for the orbit perpendicular to the axis of revolution, where $m_0$ is the mass of free electron) is much smaller than that in LaBi ($m^*_{perp}$ = $0.25m_0$) [12]. As we know, conductivity is proportional to $e^2\tau/m^*$, smaller $m^*$ will give larger conductivity. Hence, the combined conductivities of both hole pockets $\sigma_h$ and the third electron pocket $\sigma_e$(xx) in LaSb is larger than those in LaBi, thereby decreasing the polarization in LaSb.

The value of polarization in this class of materials can further be enhanced by diminishing the conductivity from the three dimensional hole pockets. One possibility would be to electron dope the parent material to minimize the volume of the hole Fermi surface. However, maintaining the large MR in such cases would be challenging because the large MR is believed to the outcome of the carrier compensation in this class of compounds. This would be an interesting possibility to explore in the future.

## Conclusions

We observed a large valley polarization of electrons in the topological semimetal LaBi under magnetic field. Because of the large mass anisotropy, a valley polarization as large as 60 % could be achieved despite the presence of additional three-dimensional hole pockets. The field-dependent anisotropy factor behaves differently compared to bismuth and $SrMnBi_2$. LaBi is a truly three-dimensional compound and can inspire further research for low dimensional Fermi surfaces and valleytronics in 3D systems. The sister compound LaSb exhibits a weaker polarization despite a larger mass anisotropy. Our work on LaBi and the related compounds is an important addition to a very few compounds where the valley polarization can be manipulated using rotating magnetic field.

## Acknowledgements

This work was financially supported by the ERC Advanced Grant No. (742068) 'TOPMAT'. We thank K. Götze and Dr. Shu-Chun Wu for valuable discussions. We acknowledge the support of the High Magnetic Field Laboratory Dresden (HLD) at HZDR, member of the European Magnetic Field Laboratory (EMFL).

Figure Captions

FIG. 1. (a) Crystal structure of LaBi showing rock-salt-type atomic arrangement of atoms. (b) Fermi surfaces (FS) in the first Brillouin zone, where highly elongated cigar-like pockets (grey) are centered at the X-points. Two hole pockets (dark brown and light brown) centered at the $\Gamma$-point are comparatively isotropic. (c) Schematic of the transport measurement on the LaBi crystal where current is applied along the *a*-axis and the magnetic field is rotated in the *bc*-plane. The geometry guarantees a constant contribution of the Lorentz force throughout the measurement. (d) Anisotropy of the electron pockets depicted by showing the experimentally determined extremal FS area as a function of $\phi$. The area is calculated from the SdH quantum-oscillation frequencies by use of the Onsager relation: $F = \left(\Phi_0/2\pi^2\right) A_k$, where $\Phi_0$ (2.07 × 10$^{-15}$ Tm$^2$) is the flux quantum, and $A_k$ is the extremal FS area.

FIG. 2. (a) Resistivity of LaBi as a function of $\phi$ at different applied magnetic fields. The effect of the anisotropic electron valleys on the resistivity is present already at a field as small as 0.3 T and the effect increases with increasing magnetic field. Polar plots of the conductivity at 9 T for (b) 2 K and (c) 100 K.

FIG. 3. (a) Conductivity of LaBi at 2 K and 9 T as a function of $\phi$. The solid red line is a fit to the conductivity data according to equation 3. Solid green and blue lines are the contributions from valleys along $k_y$ and $k_z$, respectively. The dotted orange line is the contribution from the two hole pockets and the electron pockets aligned in the $k_z$ direction. (b) Corresponding total polarization (solid red line) of electrons with contributions from even (green line) and odd (blue line) electron valleys.

FIG. 4. (a) Anisotropy factor *r* as a function of magnetic field at 2 and 30 K, calculated from the fit using equation 3. (b) Magnetic-field dependent resistivity of LaBi at 2 K for $\phi$ = 0° and 45°. (**c**) Percental increase in the resistivity at 45° compared to 0° as a function of field for different temperatures.

FIG. 5. (a) Polar plots of the electron polarizations comparing LaBi and the sister compound LaSb. (**b**) Contour plot of oscillation amplitudes in the LaSb $\phi$-frequency plane, showing a large anisotropy of the electron pockets whereas the hole pocket remains almost isotropic.

References

[1] A. Fert, Rev. Mod. Phys. **80**, 1517, (2008).
[2] I. Žutić, J. Fabian, and S. Das Sarma, Rev. Mod. Phys. **76**, 323, (2004).
[3] S.-W. Cheong, and M. Mostovoy, Nat. Mater. **6**, 13, (2007).
[4] R. Ramesh, and N. A. Spaldin, Nat. Mater. **6**, 21, (2007).
[5] A. Rycerz, J. Tworzydlo, and C. W. J. Beenakker, Nat. Phys. **3**, 172, (2007).
[6] Y. Shimazaki, M. Yamamoto, I. V. Borzenets, K. Watanabe, T. Taniguchi, and S. Tarucha, Nat. Phys. **11**, 1032, (2015).
[7] H. Zeng, J. Dai, W. Yao, D. Xiao, and X. Cui, Nat. Nano. **7**, 490, (2012).
[8] K. F. Mak, K. He, J. Shan, and T. F. Heinz, Nat. Nano. **7**, 494, (2012).
[9] SuzukiR, SakanoM, Y. J. Zhang, AkashiR, MorikawaD, HarasawaA, YajiK, KurodaK, MiyamotoK, OkudaT, IshizakaK, AritaR, and IwasaY, Nat. Nano. **9**, 611, (2014).
[10] Z. Zhu, A. Collaudin, B. Fauque, W. Kang, and K. Behnia, Nat. Phys. **8**, 89, (2012).
[11] Y. J. Jo, J. Park, G. Lee, M. J. Eom, E. S. Choi, J. H. Shim, W. Kang, and J. S. Kim, Phys. Rev. Lett. **113**, 156602, (2014).
[12] A. Hasegawa, J. Phys. Soc. Jpn. **54**, 677, (1985).
[13] M. Zeng, C. Fang, G. Chang, Y.-A. Chen, T. Hsieh, A. Bansil, H. Lin, and L. Fu, arXiv:1504.03492, (2015).
[14] N. Kumar, C. Shekhar, S.-C. Wu, I. Leermakers, O. Young, U. Zeitler, B. Yan, and C. Felser, Phys. Rev. B **93**, 241106, (2016).
[15] S. Shanshan, W. Qi, G. Peng-Jie, L. Kai, and L. Hechang, New J. Phys. **18**, 082002, (2016).
[16] F. Fallah Tafti, Q. Gibson, S. Kushwaha, J. W. Krizan, N. Haldolaarachchige, and R. J. Cava, Proc. Natl. Acad. Sci. **113**, E3475, (2016).
[17] Y. Wu, T. Kong, L.-L. Wang, D. D. Johnson, D. Mou, L. Huang, B. Schrunk, S. L. Bud'ko, P. C. Canfield, and A. Kaminski, Phys. Rev. B **94**, 081108, (2016).
[18] L. K. Zeng, R. Lou, D. S. Wu, Q. N. Xu, P. J. Guo, L. Y. Kong, Y. G. Zhong, J. Z. Ma, B. B. Fu, P. Richard, P. Wang, G. T. Liu, L. Lu, Y. B. Huang, C. Fang, S. S. Sun, Q. Wang, L. Wang, Y. G. Shi, H. M. Weng, H. C. Lei, K. Liu, S. C. Wang, T. Qian, J. L. Luo, and H. Ding, Phys. Rev. Lett. **117**, 127204, (2016).
[19] J. Nayak, S.-C. Wu, N. Kumar, C. Shekhar, S. Singh, J. Fink, E. E. D. Rienks, G. H. Fecher, S. S. P. Parkin, B. Yan, and C. Felser, Nat. Commun. **8**, 13942, (2017).
[20] L.-K. Zeng, R. Lou, D.-S. Wu, P.-J. Guo, L.-Y. Kong, Y.-G. Zhong, J.-Z. Ma, B.-B. Fu, P. Richard, and P. Wang, arXiv:1604.08142, (2016).
[21] J. Isberg, M. Gabrysch, J. Hammersberg, S. Majdi, K. K. Kovi, and D. J. Twitchen, Nat. Mater. **12**, 760, (2013).
[22] F. F. Tafti, M. S. Torikachvili, R. L. Stillwell, B. Baer, E. Stavrou, S. T. Weir, Y. K. Vohra, H. Y. Yang, E. F. McDonnell, S. K. Kushwaha, Q. D. Gibson, R. J. Cava, and J. R. Jeffries, Phys. Rev. B **95**, 014507, (2017).
[23] D. A. Abanin, S. A. Parameswaran, S. A. Kivelson, and S. L. Sondhi, Phys. Rev. B **82**, 035428, (2010).
[24] B. E. Feldman, M. T. Randeria, A. Gyenis, F. Wu, H. Ji, R. J. Cava, A. H. MacDonald, and A. Yazdani, Science **354**, 316, (2016).

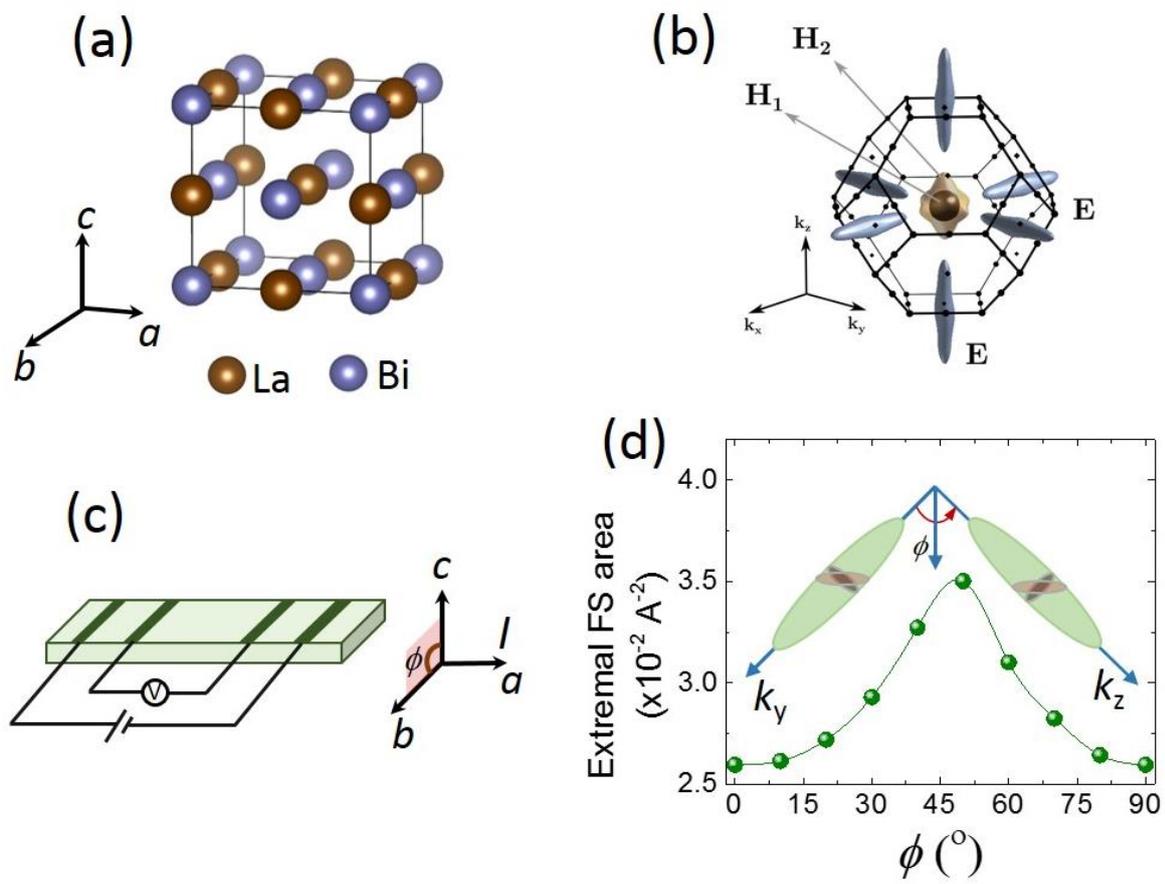

FIG. 1

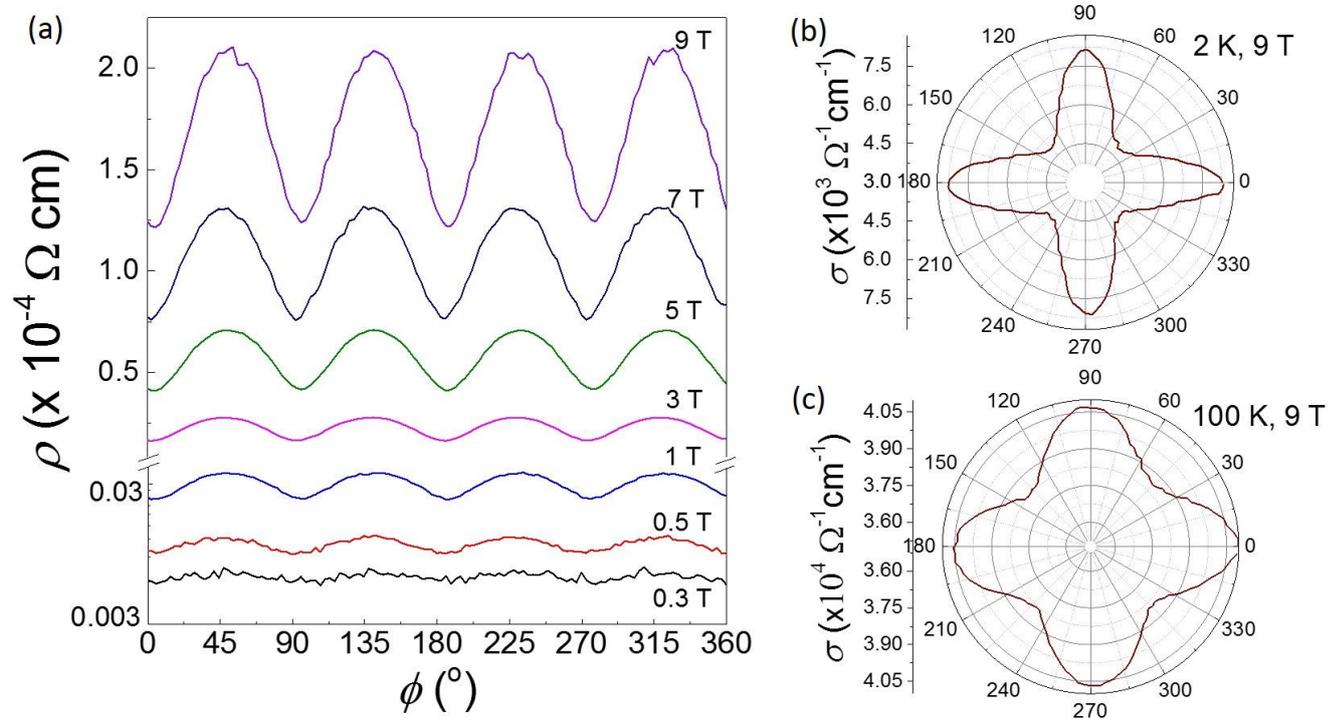

FIG. 2

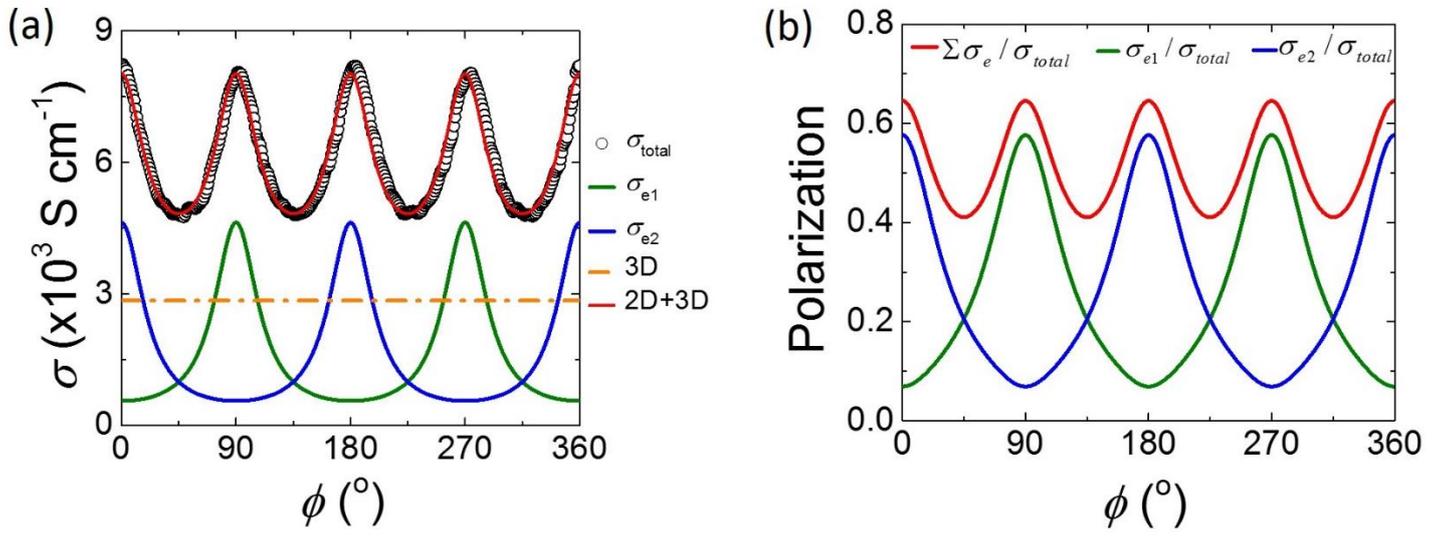

FIG. 3

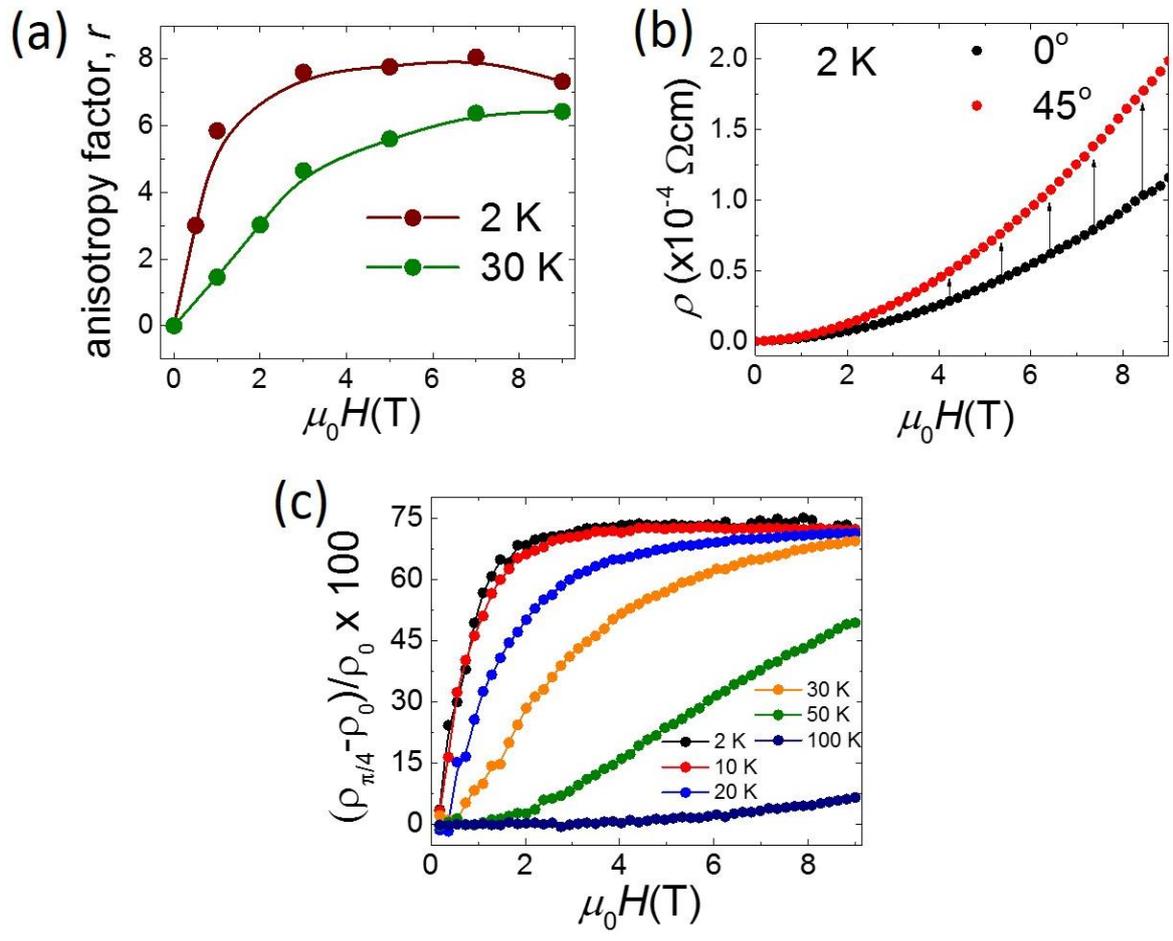

FIG. 4

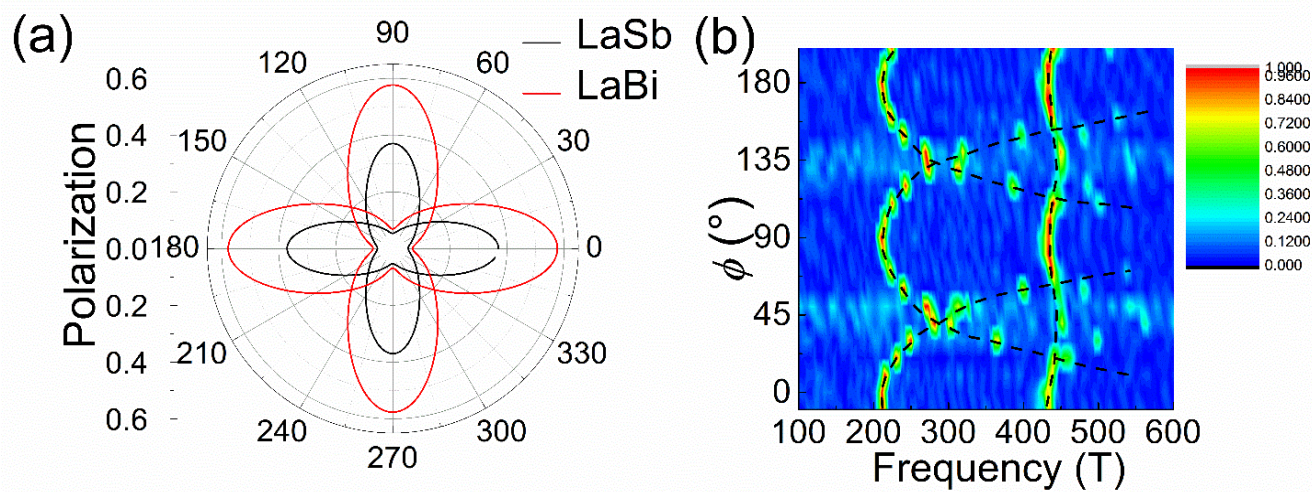

FIG. 5